\input{psfig}
\input{epsf}

\def\mev{\,{\rm Me\kern-0.1em V}}
\def\gev{\,{\rm Ge\kern-0.1em V}}

\documentstyle[11pt]{article}

\renewcommand{\baselinestretch}{1.8}
\begin{document}
\vspace*{-1.25in}
\small{
\begin{flushright}
FERMILAB-Pub-00/306-T \\[-.1in]
Nov~2000 \\
\end{flushright}}
\vspace*{.75in}
\begin{center}
{\Large{\bf  String Breaking in Four Dimensional Lattice QCD}}\\
\vspace*{.45in}
{\large{A.~Duncan$^1$, 
E.~Eichten$^2$,  and
H.~Thacker$^3$}} \\ 
\vspace*{.15in}
$^1$Dept. of Physics and Astronomy, Univ. of Pittsburgh, 
Pittsburgh, PA 15260\\
$^2$Fermilab, P.O. Box 500, Batavia, IL 60510 \\
$^3$Dept.of Physics, University of Virginia, Charlottesville, 
VA 22901
\end{center}
\vspace*{.3in}
\begin{abstract}
Virtual quark pair screening leads to breaking of the   
string between fundamental representation quarks in QCD.
For unquenched four dimensional lattice QCD,  
this (so far elusive) phenomenon is studied using the 
recently developed truncated determinant algorithm (TDA).
The dynamical configurations were generated on an Athlon 650 MHz PC.
Quark eigenmodes up to 420 MeV are included exactly in these
TDA studies performed at low quark mass on large coarse (but O($a^2$)
improved) lattices. 
A study of Wilson line correlators in Coulomb gauge extracted from an 
ensemble of 1000 two-flavor dynamical configurations reveals evidence 
for flattening of the string tension at distances \newline 
$R{\ \lower-1.2 pt\vbox{\hbox{\rlap{$>$}\lower5pt\vbox{\hbox{$\sim$}}}}\ }$ 1 fm.
\end{abstract}

\newpage
\renewcommand{\baselinestretch}{1.0}

\section{Introduction}

  The behavior of the static energy of a quark-antiquark pair at large distance
provides perhaps the most striking qualitative difference between quenched and 
full QCD. In the quenched theory virtual quark-antiquark pairs unconnected to 
external sources are neglected and the string tension rises indefinitely at large 
distance,  whereas the full theory automatically screens the quark-antiquark 
potential at  large distances by populating the vacuum with dynamical quark 
pairs. It is therefore not surprising that the demonstration of string breaking 
has long been regarded as a classic bellwether for testing the efficacy of 
dynamical QCD algorithms. Unfortunately, despite numerous studies \cite{schilling}
and the expenditure of a large amount of computational effort, the direct 
observation of string breaking in zero-temperature 4-dimensional unquenched QCD, 
in which the string tension is seen to become essentially flat at large distances,
has not yet been clearly established. On the other hand, the phenomenon has been seen 
quite convincingly in lower dimensional gauge theories such as QED2 \cite{fullqcdpaper},
QCD3 \cite{Trottier}, or in QCD4 at finite temperature \cite{9809105}. 
The expected softening (though not flattening) of the potential due to sea-quarks 
has also been seen in recent work on three flavor QCD \cite{bernard}.

  The reasons suggested for the failure to observe a clear signal of string breaking 
in zero temperature 4-dimensional QCD range from the inability to decouple the
lowest energy string states at the still rather small Euclidean time extents of the 
measured Wilson loops \cite{Trottier,9912044} to the existence of a 
completely new phase of the string in which breaking is completely 
invisible \cite{9903013}. In the former case, it has been suggested 
that use of an improved string operator which suppresses appropriately the coupling
of higher energy string states as the breaking point is approached is a prerequisite
for exposing the desired flattening of the string tension .
The results that we present in this paper confirm that string breaking only appears in Wilson line
correlators at sufficiently large Euclidean times, but demonstrate
the breaking directly at the level of unsmeared Wilson lines and without 
explicitly mixing in two-meson states \cite{2meson} in the initial string state. 

The three major differences between the simulations described here and previous 
studies of stringbreaking are
\begin{enumerate}
\item We measure the static energy of Wilson lines in Coulomb gauge rather than 
the large area behavior of Wilson loops. The Wilson line operator in Coulomb gauge
has a larger overlap with the lowest energy states of a static quark-antiquark 
source pair than the Wilson loop. (In the case of abelian gauge theory, the overlap 
for the Wilson line operator is perfect.)
\item We work on physically large coarse lattices, which allows us to go to large
Euclidean time (up to 1.2 fm) in order to project out the lowest state, which for large
distance corresponds to a meson-antimeson pair.
\item We work at relatively small quark mass (pion mass about 195 MeV). As emphasized 
previously \cite{fullqcdpaper}, the truncated determinant algorithm works perfectly
well at arbitrarily small quark masses, as the convergence rate of the Lanczos algorithm
used to extract the low eigenvalues depends only on the density of  the infrared
spectrum and does not dramatically deteriorate as we lower the quark mass. 
\end{enumerate}

  The dynamical fermion algorithm used in this paper has been described in 
considerable detail elsewhere \cite{fullqcdpaper} so we will merely summarize
the basic features. The hermitian quark Dirac operator has a completely gauge invariant
spectrum which can therefore be gauge-invariantly split into a set of infrared
modes (up to some momentum cutoff large enough to encompass the desired infrared
virtual quark physics), while the ultraviolet modes are very accurately modelled
by a local gauge-invariant pure-gauge action \cite{tonypisa} which for large distance 
physics mainly results in a coupling renormalization.  In the truncated determinant
algorithm (TDA), we perform a simulation of  the theory including {\em exactly}
all the infrared eigenvalues in the quark determinant up to the chosen cutoff (for a
detailed description of this approach see \cite{fullqcdpaper}). The
Lanczos technique used to extract the low eigenvalues does not suffer from 
rapidly increasing convergence time even at very low quark masses (in contrast to HMC algorithms,
where the quark inversions become prohibitively expensive in this limit), so we are
able to work essentially at kappa critical. Of course, string breaking is expected to
set in earlier for light dynamical quarks, so the TDA approach has a natural advantage
over other dynamical QCD schemes for this problem. The other important feature of
the calculations described in this paper is the use of large coarse lattices with an
O($a^2$) improved gauge action to restore the rotational invariance of the
measured static energies. Specifically, we give results for simulations performed
on 6$^4$ lattices with the improved action $S=$3.7[$1.0(\rm plaq) +1.04(\rm trt)$],
using the notation of \cite{alfordetal}. 
Although there is considerable scale uncertainty on such a course lattice,
we estimate the lattice spacing for this theory is about 0.4 fm.
We have performed the simulations at $\kappa$=0.2050 corresponding to 
$m_{\pi} \simeq$ 195 MeV.  
Preliminary results from these simulations have been reported earlier \cite{estipisa}.

  In Section 2 we describe the computational and statistical issues underlying our
results. Some further details of the computational load required by the
TDA method are discussed. 
We describe the equilibration of our configurations in the TDA
simulations, and the autocorrelation data underlying our error analysis are given.
In Section 3 results for the correlators of Wilson lines at Euclidean times 
0.4, 0.8, and 1.2 fm (1,2 and 3 lattice spacings) are reported. 
The flattening of the static energy associated with string breaking is finally
visible for T$ \simeq$ 1 fm. 
In Section 4 we summarize our conclusions and indicate ongoing calculations.

\newpage

\section{TDA Simulations on large coarse lattices}

  In the TDA approach to dynamical QCD, the quark determinant
${\cal D}(A)= $ det$(\gamma_{5}(D\!\!\!/(A)-m)) \equiv $ det$(H)$ 
is gauge-invariantly split into infrared and ultraviolet parts
\begin{equation}
 {\cal D}(A) = {\cal D}_{IR}(A){\cal D}_{UV}(A)
\end{equation}
by introducing a cutoff $\Lambda_{\rm cut}$ on the absolute magnitude of the 
eigenvalues $\lambda_{i}$ of the hermitian Dirac operator $H$. These eigenvalues 
are gauge-invariant  generalizations of  the quark offshellness 
(i.e. for $A\rightarrow 0$, $\lambda_{i}\rightarrow \pm \sqrt{p^2+m^2}$)) and the 
cutoff is chosen to include as much as possible of the low energy structure of 
the unquenched theory while leaving the fluctuations of ln${\cal D}_{IR}(A)$ 
(which is included {\em exactly} in the Boltzmann measure of the simulation) of 
order unity after each sweep updating all gauge links.  Fortunately, this choice is
possible on lattices of large physical size as well as for light quark masses 
close to the critical value. In the simulations reported
in this paper, the cutoff $\Lambda_{\rm cut}$ is chosen at about 420 MeV. On the lattices 
generated this corresponds to including the lowest 840 eigenvalues of $H$. 
These eigenvalues are extracted by a Lanczos procedure for each trial gauge configuration
generated using an improved gauge action and the new configuration is then
subjected to a Metropolis accept/reject step based on the change in 
$N_{\rm flav}$ln${\cal D}_{IR}(A)$ (we have used $N_{\rm flav}$=2 
flavors of degenerate light  quarks in the simulations).
Detailed balance in this procedure is ensured by using a random link choice 
procedure in the pure gauge update step.  We have stored gauge configurations 
after ten combined gauge-update + metropolis determinant accept/reject steps.
(The metropolis step had a typical acceptance rate of $50\%$.)

  The computational load in these simulations is completely dominated by the extraction 
of the infrared quark eigenmodes (typically to seven or eight place accuracy). 
For example, on a 6$^4$ lattice, the gauge update takes a
few seconds while the calculation of  ${\cal D}_{IR}$ takes about 9 minutes on an
Athlon 650-MHz processor. The string breaking results reported  in Section 3
were performed on such a processor and required about 2.0 processor-months to 
accumulate 1000 configurations once equilibration is reached.

   As mentioned previously, we work on coarse 6$^4$ lattices but with O($a^2$) improved
gauge action. Following Alford et al \cite{alfordetal}, we improve the gauge action with
a single additional operator, with coefficients tuned to optimize rotational invariance
of the string tension
\begin{eqnarray}
  S(U) &=& \beta_{\rm plaq}\sum_{\rm plaq}\frac{1}{3}{\rm ReTr}(1-U_{\rm plaq}) \nonumber
\\
            &+& \beta_{\rm trt}\sum_{\rm trt}\frac{1}{3}{\rm ReTr}(1-U_{\rm trt})
\end{eqnarray}
where ``trt"  refers to a 8 link loop of generic structure (+x,+y,+x,-y,-x,+y,-x,-y) 
(the ``twisted rectangle" of Ref\cite{alfordetal}).
With the choices $\beta_{\rm plaq}$=3.7,
$\beta_{\rm trt}$=1.04$\beta_{\rm plaq}$,  the quenched static
quark potential becomes a smooth function of lattice radial separation \cite{alfordetal}
even on these very
coarse lattices, with lattice spacing $a\simeq$ 0.4 fm (for the unquenched theory). As we
do not improve the quark action, the lattice spacing quoted here is determined by
matching the initial linear rise of the string tension to a physical value.
The results given below show
that the restoration of rotational invariance survives reasonably well
the introduction of the quark determinant, 
so that we have not found it necessary to retune the pure gauge action.

\begin{figure}
\psfig{figure=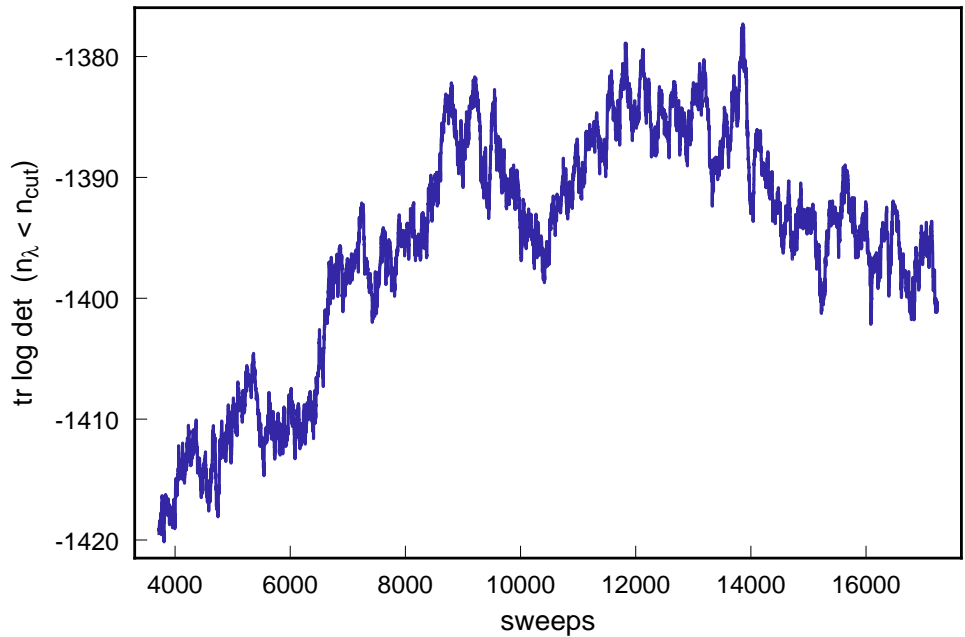,
width=0.95\hsize}
\caption{Equilibration of the quark determinant in TDA simulation of a coarse 6$^4$ lattice}
\label{fig:equildet}
\end{figure}

  To maximize our chances of seeing string breaking within the spatial limitations of the
lattices being used we have chosen a kappa value corresponding to a rather light pion, 
namely $\kappa$=0.2050, corresponding to a pion mass of about 195 MeV (or 0.39 in
lattice units).  This does not seem to result in a serious loss of acceptance at the 
level of individual Monte Carlo steps, but the equilibration process (starting from a 
quenched initial configuration) is definitely
slower in comparison to TDA simulations performed with heavier quarks on lattices
of smaller physical size \cite{fullqcdpaper}. The sequence of infrared determinant
values (specifically, ln${\cal D}_{IR}(A)$) generated in a TDA simulation starting
from a quenched configuration is shown in Fig.1. It appears that about 7000 Monte Carlo
sweeps were needed to reach equilibrium  
(corresponding to about 1.5 650-MHz Athlon processor months). 
The results reported in the next section were based on the 1000 saved configurations 
between sweeps 7000 and 169900 (see Figure \ref{fig:equildet}).

\begin{figure}
\psfig{figure=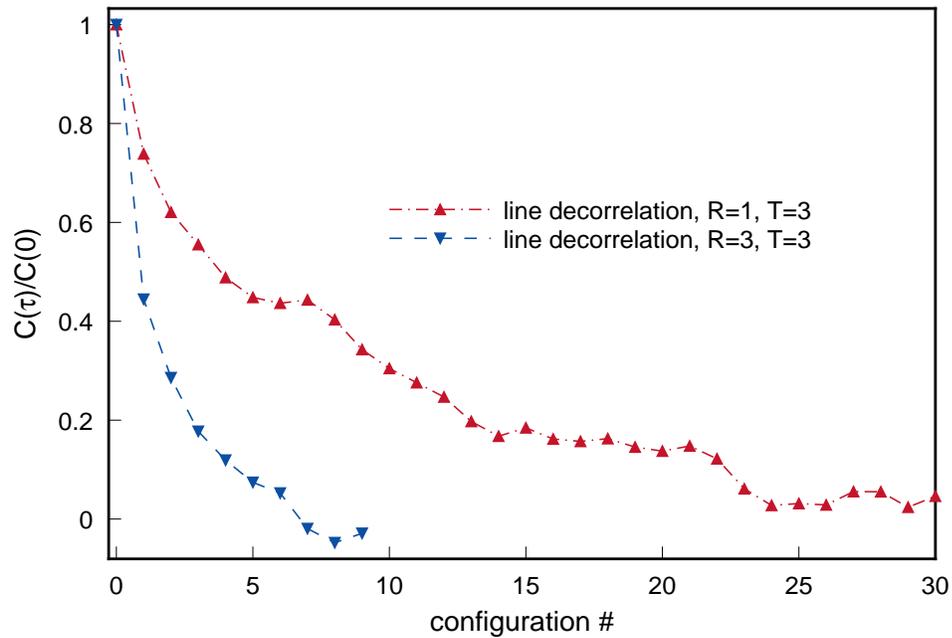,
width=0.95\hsize}
\caption{Typical autocorrelation curves of Wilson line correlators (R=1,3; T=3)}
\label{fig:autocorr}
\end{figure}

  Given the rather long times needed for equilibration, the issue of autocorrelation of
measured quantities in the post-equilibrium  configurations naturally assumes
great importance. We have performed a careful study of the decorrelation rate of all
the Wilson line correlators used to extract the static energy plots given in Section 3
below. Thus, if $W(\tau)$ is a Wilson line correlator measured for the $\tau$'th saved
configuration, an autocorrelation function can be computed as the ensemble average
\begin{equation}
  C(\tau) \equiv \sum_{n}(W(n+\tau)W(n)-<W>^{2})
\end{equation}
The autocorrelation time can then be read off from the exponential decay of $C(\tau)$.
Typical autocorrelation curves are shown in Figure \ref{fig:autocorr}, 
which shows the decorrelation of 
Wilson line correlators spatially separated by 1 or 3 lattice units and of Euclidean time
extent 3, measured on an ensemble of 300 successive saved configurations.
For $R=1$, an exponential fit gives an autocorrelation time of 7.9 
(in units of 10 Monte Carlo steps), while the area
under the autocorrelation curve gives 8.4 for the same quantity.  For $R=3$, the autocorrelation
time is of order unity (1.5, from an exponential fit).  
In general, autocorrelation times range from about unity for the largest loops 
to on the order of 20 for the smallest. 
We have therefore assumed that line correlators from successive bins of 20 (or more)
 saved
configurations are effectively decorrelated for all relevant $R$ and $T$. 

\begin{figure}
\psfig{figure=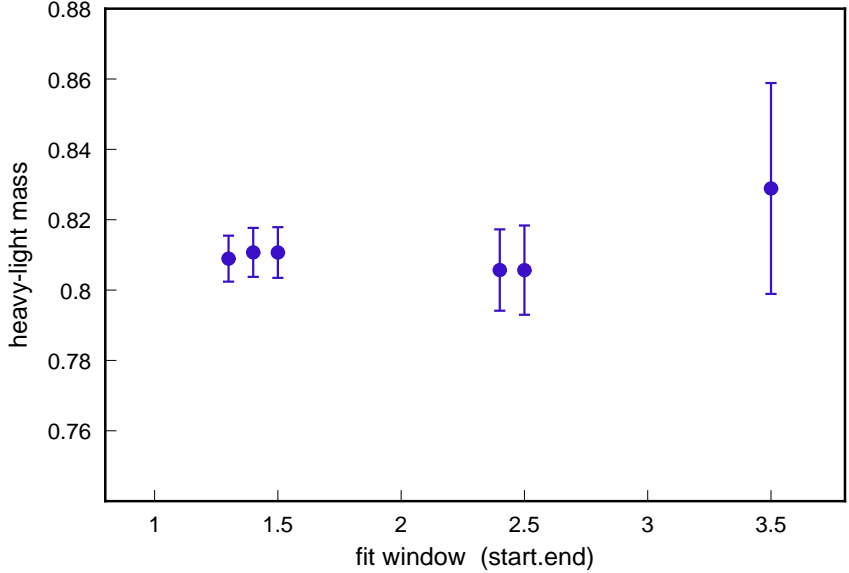,
width=0.85\hsize}
\caption{Various fits to ground state heavy-light meson mass. The start time and end time
are represented as the decimal start.end}.
\label{fig:hl}
\end{figure}

 We have calculated the standard light meson and baryon spectrum, the eta prime,
and the heavy-light mesons. We will report on these results in a future 
paper. The ground state heavy light meson mass is relevant to the static energy analysis.
As shown in Figure \ref{fig:hl}, a consistent mass value was obtained for all the 
various fitting time windows. 
The mass fit using $T=1-5$ was $m_{HL} = 0.811 \pm .007$. 
\newpage

\section{Static Energy Results}

  The static energy of a quark-antiquark pair is calculated using the
Euclidean time evolution of a color-singlet pair in Coulomb gauge. The
relevant correlator is
\begin{equation}
  W(\vec{R},T) = <\bar{\Psi}(0,T)\Psi(\vec{R},T)\bar{\Psi}(\vec{R},0)\Psi(0,0)>_{\rm coul}
\end{equation}
where $\Psi$ is an infinitely massive quark field and the correlator is evaluated in
Coulomb gauge. The static energy $V(\vec{R})$ is then defined as the energy of the
lowest state coupled by $\bar{\Psi}(\vec{R},0)\Psi(0,0)$ to the vacuum, i.e.
\begin{equation}
  W(\vec{R},T) \rightarrow C\exp{(-TV(\vec{R}))},\;\;\;T\rightarrow\infty
\end{equation}
or equivalently
\begin{equation}
  V(\vec{R}) \equiv \lim_{T\rightarrow\infty}\ln{\frac{W(\vec{R},T-1)}{W(\vec{R},T)}}
\end{equation}
On the lattice the correlator (4) is evaluated  as an ensemble average of
${\rm Tr}(L(0,T)L^{\dagger}(\vec{R},T))$, where $L(\vec{R},T)$ is a Wilson line-
i.e. a product of $T$ adjacent link matrices in the temporal direction at spatial location 
$\vec{R}$. As mentioned previously, there is a distinct advantage to the use of
Coulomb gauge Wilson line correlators over Wilson loop expectations in the
study of  stringbreaking effects. The Wilson loop necessarily
involves additional contributions from intermediate
states containing transverse gluons which are absent in the Coulomb gauge
Wilson line correlators. In perturbation theory these states correspond to diagrams
in which gluons are exchanged between the top and bottom horizontal (i.e. fixed time) portions
of the Wilson loop. In quenched abelian theory, for example, the correlator (4) is
a pure exponential $  W(\vec{R},T) \propto e^{-V(R)T}$, whereas the corresponding
Wilson loop expectation is proportional to  $e^{-(V(R)T+V(T)R)}$ implying the presence of
excited states. It is therefore reasonable to expect that stringbreaking in unquenched
QCD will emerge more
rapidly (i.e. at smaller Euclidean times $T$) if the Coulomb gauge correlator (4) is used.

  Since the calculation is done on symmetrical 6$^4$ lattices, 
we actually obtain four sets of correlators
for each gauge configuration, obtained by successively gaugefixing to Coulomb gauge
with the time direction chosen as each of the four original Euclidean spacetime
directions. Exploiting this degeneracy provides a valuable increase in the statistics,
as we effectively have an ensemble of 4000 correlators from the original set of
1000 gauge configurations. The autocorrelation times described in the preceding
section were obtained however by preaveraging the four correlator sets coming from each
configuration.

\begin{figure}
\psfig{figure=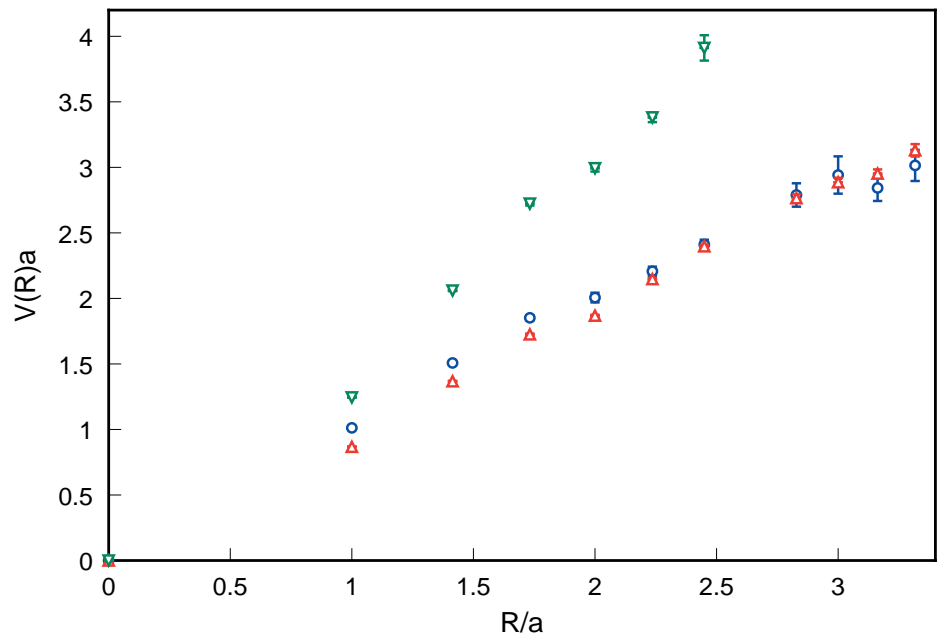,
width=0.95\hsize}
\caption{Static energy from Wilson line correlator (T=2) for 
unquenched QCD and quenched (untuned and tuned) QCD.}
\label{fig:pot2}
\end{figure}

\begin{figure}
\psfig{figure=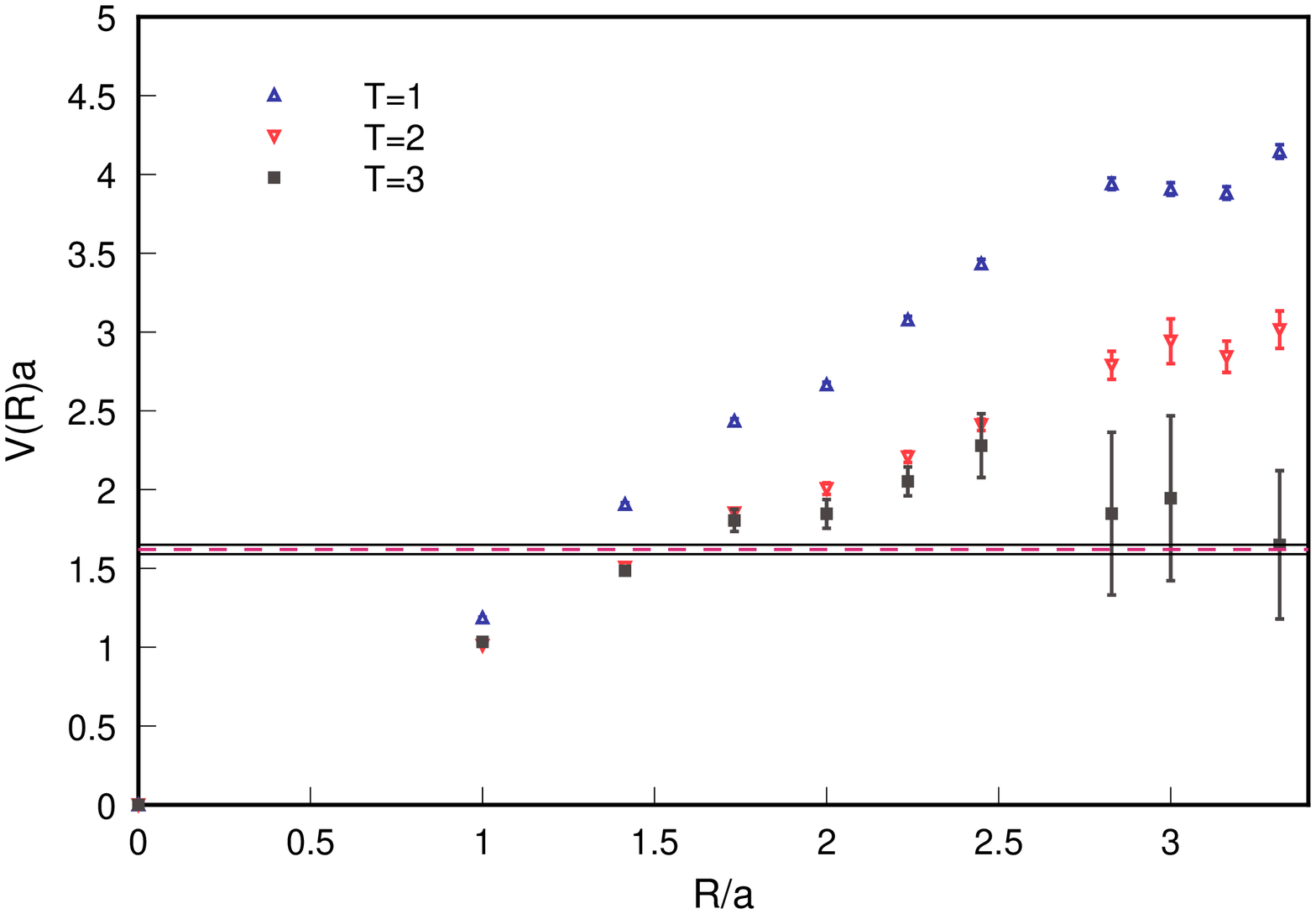,
width=0.95\hsize}
\caption{Static energy from unquenched QCD Wilson line correlators (T=1,2,3). The 
calculated heavy-light meson pair threshhold is also indicated.}
\label{fig:pot1-3}
\end{figure}

In Figure \ref{fig:pot2} the static energy at $T=2$ (circles) is compared with
the similar results for two other theories.  First, we show (down triangles)
the $T=2$ static energy for 
the quenched theory (with identical improved gauge action (2) but with the
determinant contribution switched off), obtained from an ensemble of 8000 configurations,
but otherwise analysed identically to the dynamical configurations (the untuned quenched
theory).  Both a shift in scale at short distances and the deviation at longer distances 
are clearly observed.  Second, the static energy at $T=2$ for the original Alford et. al. 
\cite{alfordetal} unquenched action is plotted (up triangles). This curve was used to tune our action to obtain 
an approximately equal initial slope. 
 
In Figure \ref{fig:pot1-3} we show the static energy curves for $T$=1,2 and 3 in (6) 
respectively. For comparison the heavy-light meson pair production threshhold is also
shown.
As we are working on a large coarse lattice (lattice spacing $a\simeq$0.4 fm) 3 temporal lattice
spacings already represents a fairly large Euclidean time $T\simeq 1.2$ fm and the 
signal to noise ratio at larger spatial distances has clearly deteriorated 
substantially at $T$=3. The shortest time
evolution, $T$=1 shown in Fig(3), on the other hand is still contaminated by higher energy
states and the flattening of the string tension is certainly not visible.  
However, at $T$=3  the  string tension appears to flatten out 
convincingly for distances 
$R{\ \lower-1.2pt\vbox{\hbox{\rlap{$>$}\lower5pt\vbox{\hbox{$\sim$}}}}\ }$ 1 fm (=2.5$a$). 
(Further simulations are in progress to substantially reduce the statistical errors in this
regime.)
Moreover, the smoothness of the potential
curve within the statistical errors suggests that the improved action terms are doing a
good job of restoring rotational invariance on this very coarse lattice. 

   The errors in the static potential are obtained most simply by
computing a separate error on the line correlators $W(\vec{R},T)$ and $W(\vec{R},T-1)$
in (6), using the measured standard deviation and an autocorrelation time extracted
separately for every loop size $R,T$. The errors for the log ratio in (6) can then be
obtained by combining the numerator and denominator errors in quadrature.  This
approach however almost surely yields
an overestimate of the true errors, as $W(\vec{R},T)$ and $W(\vec{R},T-1)$
tend to be {\em positively} correlated, reducing the variance in the ratio.  Line
correlators extracted from individual configurations on a small lattice tend to be 
extremely noisy (especially for the large loops of interest here) so to examine 
this correlation we have performed the error analysis by binning the 1000 configurations 
into 40 sets of 25 consecutive configurations. 
The average line correlators in each bin are then 
essentially decorrelated and a straightforward jackknife analysis can be performed on
the ratios in (6). This is the approach used to obtain the errors 
displayed in Figure \ref{fig:pot1-3}.
\newpage
\section{Conclusions}

  The analysis of dynamical two-flavor configurations on large coarse lattices obtained by the
truncated determinant simulation method provides evidence for string breaking
in zero temperature 4-dimensional QCD. One clear advantage of the method essential for
its success in this case is the ability to generate a sufficient number of equilibrated
and decorrelated dynamical configurations even for light quark mass on a physically large 
lattice, where standard HMC simulations would encounter computational problems. 
Evidently, as is the case for QED2 \cite{fullqcdpaper}, the infrared quark modes 
included in the determinant in the TDA contain all the essential physics of string breaking.

Simulations on larger coarse lattices (8$^4$ lattices with the same action (2)) as
well as at different kappa values for the two light quark flavors are in progress. 
In particular, we hope to study in detail the hairpin amplitudes relevant to the 
eta-prime mass. Here also, the essential physics should be obtainable from amplitudes 
measured in a TDA simulation \cite{estipisa}. 
 
\newpage
\section{Acknowledgements}

The work of A. Duncan was supported in part by NSF grant PHY00-88946.
The work of  E. Eichten was performed at the Fermi National
Accelerator Laboratory, which is operated by University Research Association,
Inc., under contract DE-AC02-76CHO3000. 
The work of H. Thacker was supported in part by the Department of Energy
under grant DE-FG02-97ER41027. The numerical work was performed on a
9 node Beowulf system at the University of Pittsburgh. 



\end{document}